# On graph theoretic results underlying the analysis of consensus in multi-agent systems

Pavel Chebotarev[1]

Institute of Control Sciences of the Russian Academy of Sciences
65 Profsoyuznaya Street, Moscow 117997, Russia

The objective of this note is to give several comments regarding the paper [1] published in the Proceedings of the IEEE.

As stated in the Introduction of [1], "*Graph Laplacians* and their spectral properties […] are important graph-related matrices that play a crucial role in convergence analysis of consensus and alignment algorithms." In particular, the stability properties of the distributed consensus algorithms

$$\dot{x}_i(t) = \sum_{j \in N_i} a_{ij}(t)(x_j(t) - x_i(t)), \quad i = 1,\ldots,n \qquad (1)$$

for networked multi-agent systems are completely determined by the location of the Laplacian eigenvalues of the network. The convergence analysis of such systems is based on the following lemma [1, p. 221]:

*Lemma 2*: (spectral localization) Let $G$ be a strongly connected digraph on $n$ nodes. Then $\mathrm{rank}(L) = n - 1$ and all nontrivial eigenvalues of $L$ have positive real parts. Furthermore, suppose $G$ has $c \geq 1$ strongly connected components, then $\mathrm{rank}(L) = n - c$.

Here, $L = [l_{ij}]$ is the Laplacian matrix of $G$, i.e., $L = D - A$, where $A$ is the adjacency matrix of $G$, and $D$ is the diagonal matrix of vertex out-degrees.

Four comments need to be made concerning this lemma.

First, the last statement of the lemma is not correct. Indeed, recall that the strongly connected components (SCC) of a digraph $G$ are its maximal strongly connected subgraphs. For instance, if $G$ is *a converging tree*, i.e., $G$ is a directed tree with root $r$ such that every vertex of $G$ can be linked to $r$ via a directed path, and $n > 1$, then $G$ has $c = n$ strongly connected components, but $\mathrm{rank}(L) = n - 1 > n - c = 0$.

The statement under consideration becomes valid if one replaces strongly connected components with weakly connected components (WCC) and additionally requires that these WCC's are strong. A *weakly connected component* of $G$ is a maximal subgraph of $G$ whose vertices are mutually reachable by violating the edge directions. A more general correct statement results by substituting, in the same place, sink SCC's, where a *sink strongly connected component* is an SCC having no edges directed outwards. This result was proved in [2] as well as some other Laplacian related results applicable to the cooperative control.

Second, the proof of the rank property (the first statement of Lemma 2) is attributed in [1] to [3]. Let me note that a stronger fact was proved earlier in [2]. More specifically, Proposition 11 of [2] states that $\mathrm{rank}(L) = n - d$, where $d$ is the so-called *in-forest dimension* of $G$, i.e., the minimum possible number of converging trees in a spanning converging forest of $G$. It was also shown (Proposition 6) that the in-forest dimension of $G$ is equal to the number of its sink SCC's and that the forest dimension of a strongly connected digraph is one (Proposition 7). Consequently, for a strongly connected digraph, $\mathrm{rank}(L) = n - 1$, which coincides with the first statement of Lemma 2. In addition, according to Proposition 8, "the forest dimension of a digraph is no less than its number of weak components[2] and does not exceed the number of its strong components and the number of its unilateral components."

---

[1] E-mail: chv@member.ams.org; pavel4e@gmail.com.
[2] A weak component = a weakly connected component; a strong component = a strongly connected component.

Third, Remark 1 given after the proof of Lemma 2 says[3]: "Lemma 2 holds under a weaker condition of existence of a directed spanning tree for $G$. [...] This type of condition on existence of directed spanning trees have appeared in [4]–[6]." Here, by Lemma 2 the authors conceivably mean the conclusion that rank($L$) = $n - 1$. Let us observe that the existence of a directed spanning tree for $G$ implies that $d = 1$, so this statement follows from Proposition 11 of [2].

Fourth, the statement of Lemma 2 that "all nontrivial eigenvalues of $L$ have positive real parts" holds true in the general case, and not only for strongly connected digraphs or digraphs that contain directed spanning trees. This was shown in [7, Proposition 9].

In Section II.C of [1] a discrete-time counterpart of the consensus algorithm (1) is considered:

$$x_i(k+1) = x_i(k) + \varepsilon \sum_{j=1}^{n} a_{ij}\left(x_j(k) - x_i(k)\right), \quad i = 1,\ldots,n, \tag{2}$$

where $\varepsilon > 0$ is the step size. In the matrix form, (2) is represented as follows:

$$x(k+1) = Px(k), \tag{3}$$

where $P = I - \varepsilon L$ is referred to in [1] as the *Perron matrix* with parameter $\varepsilon$ of $G$.

The matrices $P = I - \varepsilon L$ were studied in [2] and [7]; in particular, (i) of Lemma 3 in [1] coincides with Proposition 12 of [2]. The asymptotic behavior of the process (3) is determined by the properties of the sequence $P, P^2, P^3, \ldots$. If the stochastic matrix $P$ is *primitive*, i.e., it has only one eigenvalue with modulus 1, then, as stated in Lemma 4 of [1], $\lim_{k \to \infty} P^k = vw^T$, where $v$ and $w$ are the right and left eigenvectors of $P$ corresponding to the eigenvalue 1, respectively, with a normalization that provides $v^T w = 1$. In the case of a general nonnegative Perron matrix $P$, the sequence $P, P^2, P^3, \ldots$ need not have a limit, so the *long-run transition matrix* $P^\infty = \lim_{m \to \infty} m^{-1} \sum_{k=1}^{m} P^k$ is considered. The matrix $P^\infty$ always exists and, by the *Markov chain tree theorem* proved in [8], [9], it coincides with the *normalized matrix $\bar{J}$ of maximal in-forests of $G$*. $\bar{J}$ is the eigenprojector of $L$; by Proposition 11 of [2] rank($\bar{J}$) = $d$, where $d$ is the in-forest dimension of $G$. The columns of $\bar{J}$ are the eigenvectors of $L$ corresponding to the eigenvalue 0; consequently, they determine the consensus trajectories of the process (1) and the flocking trajectories [10]. The elements of $\bar{J}$ were characterized in Theorems 2' and 3 of [2]. An algebraic method for calculating $\bar{J}$ was presented in [7].

As has been shown above, [2] and [7] contained a number of results on the Laplacians of directed graphs which were useful for the cooperative control of multi-agent systems. A number of additional results were presented in [11] and [12]. Some of them are surveyed in [13].

In January 2001 Alex Fax, one of the authors of [1], sent me a message, where he asked about the eigenstructure of digraph Laplacians and requested to send copies of related papers. During the subsequent correspondence, later in 2001, I sent him [2] and [7]. Recently, I was pleased to familiarize myself with [1] and to learn that our early results proved to be useful in the analysis of consensus and cooperation algorithms of decentralized control. However, I was surprised that, instead of references to [2] and [7], this article contained references to papers published several years later.

---

[3] The bibliographic references are redirected here to the list of references of this note.